\documentstyle[12pt]{article}

\begin{document}

\title
{
Ultracold Bose atoms in intense laser fields:\\
intensity- and density-dependent effects
}
 
\author
{K.V.Krutitsky
\thanks{
		 Permanent address:
		 Ulyanovsk Branch of Moscow Institute of Radio
		 Engineering
		 and Electronics of Russian Academy of Sciences,
		 P.B.9868,
		 48, Goncharov Str., Ulyanovsk 432011, Russia;
		 e-mail: kostya@spock.physik.uni-konstanz.de, ufire@mv.ru
},
K.-P.Marzlin
\thanks{e-mail: Peter.Marzlin@uni-konstanz.de}
and J.Audretsch
\thanks{e-mail: Juergen.Audretsch@uni-konstanz.de}
\\
Fachbereich Physik,
Universit\"at Konstanz, Fach M 674,\\
D-78457 Konstanz, Germany
}

\date{}

\maketitle

\begin{abstract}
Starting from the first principles of nonrelativistic QED we have derived
the system of Maxwell-Schr\"odinger equations, which can be used for theoretical 
description of atom optical phenomena at high densities of atoms and high
intensities of the laser radiation. 
The role of multiple atomic transitions between ground and excited states 
in atom optics has been investigated.
Nonlinear optical properties of
interacting Bose gas are studied: formula for the refractive index
has been derived and the polariton spectrum of a condensate interacting
with an intense laser field has been investigated.
\end{abstract}

\newpage

\section{Introduction}

In last decades the problem of interaction of the laser radiation with
ultracold atomic gases has attracted a lot of attention. With the aid
of the laser radiation one can manipulate the center-of-mass motion of
ultracold atoms and one can observe different wave phenomena with
ultracold atomic beams.
After the experimental realization of Bose-Einstein condensation,
which allows to create rather dense atomic systems, the problem of interaction
of photons with ultracold atoms has reached a new stage of development.
Acting with the laser on a dense atomic sample one can induce nonlinearities
in the behavior of the matter field caused by dynamical dipole-dipole interactions.
This provides a possibility to create
atomic solitons of different kinds~\cite{LEN94},
to change dramatically the effective scattering length of 
the condensate~\cite{length}, to create nonlinear beam splitters~\cite{ZHA94a},
vortices~\cite{vortices},
photonic band gaps and defeçt states in a condensate~\cite{PBG}.

In recent years different approaches to the description of interaction of
ultracold atoms in the field of optical radiation have been suggested and
different aspects of the phenomenon have been considered.
The properties of the laser radiation modified by atomic
dipole-dipole interactions were investigated~\cite{MOR95,RJ,RJcm} and
it was shown that they can be described
by the refractive index which is governed in the linear case,
when the light intensity is low enough,
by the Clausius-Mossotti
relation known from classical optics if we neglect quantum correlations.
If the quantum statistical correlations are taken
into account, the formula for the refractive index contains additional
terms defined by a position dependent correlation
function~\cite{MOR95,RJ,RJcm}. Nonlinear optical properties of
noninteracting Bose gas were studied as well~\cite{KH}.

The modification of the properties of the laser radiation should have a
back influence on the behavior of an ultracold atomic ensemble (for example,
the motion of atomic beam). An attempt to consider this back
influence was undertaken by several authors
\cite{LEN94,ZHA94a,RJ,LEN93,CAS95,WAL97}.
In the first works on the subject the two-body interactions were modelled
by the phenomenological contact potential~\cite{LEN93}.
Later on dynamical dipole-dipole interactions 
were taken into account within the framework of the nonrelativistic
electrodynamics~\cite{LEN94,ZHA94a,RJ,CAS95,WAL97,KBA99,KBA00,KMA01}.
However, in papers by W.Zhang and D.Walls \cite{ZHA94a}
and G.Lenz et al.\cite{LEN94}
the averaged polarization of
ultracold atomic ensemble was computed as a function of the incident
laser field, whereas it should be a function of the macroscopic or the
local
field which are different from the external laser field due to dynamical
dipole-dipole interactions. Wallis \cite{WAL97}
used the correct form of the equations for
the electromagnetic field. However, his equation for the matter field was
not consistent with the equation
for the electromagnetic field \cite{KBA99}. This problem was considered also
by Y.Castin and K.M{\o}lmer \cite{CAS95} and
J.Ruostekoski and J.Javanainen \cite{RJ}.
But the equations which are used in those
papers are very complicated because they are written down in terms
of the local field and dipole-dipole interactions are presented explicitly
in the form of the sum over dipole fields. This makes the analysis
very difficult and the equations, which govern the time evolution
of the matter fields, remained unsolved.

In the papers~\cite{KBA99,KBA00,KMA01} a self-consistent quantum
theory of atom optical processes has been developed.
Making use of the Lorentz-Lorenz relation, which allows to simplify 
the analysis, it was
obtained the general system of Maxwell-Schr\"odinger equations for
atomic creation and annihilation operators and the propagation equation
for the laser field which can be used for the description
of linear and nonlinear phenomena in atom optics of single- and
multi-species condensates at high
densities of the atomic system.
However, the treatment in~\cite{KBA99,KBA00,KMA01} was restricted by low
light intensities.

In the present paper we shall continue the investigations started
in~\cite{KBA99,KBA00}. Having in mind mainly atom optical applications,
we shall derive the system of Maxwell-Schr\"odinger equations for the case
of high density of the atoms and high light intensity. 
Nonlinear optical properties of the interacting
Bose gas will be also discussed.

\section{Heisenberg equations of motion for the atomic operators}

We consider a system of ultracold two-level atoms with masses $m$, 
transition frequencies $\omega_a$, and transition dipole moments ${\bf d}$.
We shall describe such a system in terms of matter field operators. 
Let $|g\rangle$ and $|e\rangle$ are the vectors of the ground and
excited states of the quantized atomic fields. Then the corresponding
annihilation operators of the atoms in these internal states are
$\hat \phi_g$ and $\hat \phi_e$.
Matter-field operators are assumed to satisfy the bosonic equal time
commutation relations.

The Heisenberg equations of motion for the atomic operators
are derived from the Hamiltonian
of the second quantized atomic field interacting with the
photons.
In the reference frame rotating with the frequency $\omega_L = c k_L$
of the incident laser field ${\bf E}_{in}({\bf r},t)$,
which is assumed to be monochromatic, and in making use of
the electric dipole approximation and
the rotating-wave approximation we obtain the following dynamical equations
for the matter-field operators~\cite{KBA99,KBA00}
\begin{eqnarray}
i \hbar
\frac{\partial \hat \phi_{g}}{\partial t}
&=&
\hat H_{cm}
\hat \phi_{g}
+
\hat H_{ge}
\hat \phi_{e}
\;,
\label{nonlinear-g}
\\
i \hbar
\frac{\partial\hat{\phi}_{e}}{\partial t}
&=&
\hat H_{cm}
\hat{\phi}_{e}
-
\hbar
\left(
\Delta + i \gamma/2
\right)
\hat{\phi}_{e}
+
\hat H_{eg}
\hat{\phi}_{g}
\;,
\label{nonlinear-e}
\end{eqnarray}
where $\hat H_{cm}=-\hbar^2 \nabla^2/(2m)$, 
$\Delta=\omega_L-\omega_a-\delta$ is the detuning of the frequency
of the laser wave from the frequency of the atomic transition,
$\delta$ and $\gamma$ are the Lamb shift and the spontaneous
emission rate of a single atom in free space, respectively.
Here we have introduced the operators 
$\hat H_{eg} = - {\bf d} \hat{{\bf E}}_{loc}^+$
and
$\hat H_{ge} = - {\bf d} \hat{{\bf E}}_{loc}^-$
which are responsible for the transitions
$|g\rangle \to |e\rangle$ and $|e\rangle \to |g\rangle$,
respectively. $\hat H_{eg}$ and $\hat H_{ge}$ are related to
the operator of the local electric field
$\hat{{\bf E}}_{loc}^\pm({\bf r},t)$. The positive-frequency part
of this operator has the form
\begin{eqnarray}
\label{def-e-local}
\hat{{\bf E}}_{loc}^+({\bf r},t)
&=&
{\bf E}_{in}^+({\bf r})
+
i
\sum_{{\bf k}\lambda}
\sqrt{\frac{2\pi\hbar\omega_k}{V}}
{\bf e}_\lambda
\hat{c}_{{\bf k}\lambda}(0)
\exp
\left[
     i {\bf k} {\bf r} 
     - 
     i
     \left( 
          \omega_k - \omega_L
     \right)
     t
\right]
\nonumber\\
&&
+
\int
d{\bf r}'
\nabla \times \nabla \times
\frac { \hat{\bf P}^+ \left( {\bf r}',t-R/c \right)}{R}
e^{
    i k_L R
  }
\;,
\end{eqnarray}
where $\nabla\times$ refers to the point ${\bf r}$ and
the operator
$\hat{c}_{{\bf k}\lambda}(0)$ corresponds to the free-space
photon field (vacuum fluctuations).
The polarization operator 
$\hat {\bf P}^+ = {\bf d}\hat\phi_{g}^\dagger \hat\phi_{e}$.
Note that in equation (\ref{def-e-local}) a small volume around the
observation point ${\bf r}$ is excluded from the integration.
The last term in equation (\ref{def-e-local}) describes dipole-dipole
interaction. Due to the structure of
the local field given by equation (\ref{def-e-local}) one can distinguish
two kinds of the photons: primary laser photons and secondary
photons, re-radiated by the atoms.

Usually in atom optics of BEC one deals with equations for the
ground state matter-field operator $\hat\phi_{g}$. Therefore,
one has to eliminate the excited state matter-field operator
$\hat\phi_{e}$ from the system of equations (\ref{nonlinear-g}),
(\ref{nonlinear-e}). In the next section we shall develop 
a general procedure of the elimination of the excited state
which is in contrast to previous works 
(see, for instance, \cite{ZHA94a}) valid for high densities
of the atoms and high intensities of the laser radiation.

\section{Elimination of the excited state}

Let's assume that initially there are no atoms in the excited state.
Then we can rewrite equation (\ref{nonlinear-e}) in the form
\begin{equation}
\hat{\phi}_{e}(t)
=
-i
\int_0^t
e^
{
  i 
  \left(
      \frac
      {
        \hat H_{cm}
      }
      {\hbar}
      - 
      \tilde\Delta
  \right)
  t'
}
\hat H_{eg}(t')
e^
{
  - i 
  \frac
  {
    \hat H_{cm}
  }
  {\hbar}
  t'
}
\hat{\phi}_{g}(t')
dt'
\;,
\end{equation}
where $\tilde\Delta = \Delta + i \gamma/2$.
Making use of the identity
\begin{equation}
\int
e^
{
  a x
}
F(x)
dx
=
\frac
{
  e^
  {
    a x
  }
}
{a}
\sum_{k=0}^{\infty}
\frac
{(-1)^k}
{a^k}
\frac{\partial^k}{\partial x^k}
F(x)
\;,
\end{equation}
which is valid for a non-vanishing $a$,
we get
\begin{equation}
\label{t-dep}
\hat{\phi}_{e}({\bf r},t)
=
\frac
{
  e^{- i \frac{\hat H_{cm}}{\hbar} t}
}
{\hbar\tilde\Delta}
\sum_{k=0}^\infty
\frac{(-i)^k}{\tilde\Delta^k}
\frac{\partial^k}{\partial t^k}
\left[
e^{i \frac{\hat H_{cm}}{\hbar} t}
\hat H_{eg}({\bf r},t)
\hat \phi_g({\bf r},t)
\right]
\;.
\end{equation}
If $\hat H_{eg}$ does not depend on time and 
the center-of-mass motion does not give a large contribution,
we have
\begin{equation}
\label{t-ind}
\hat{\phi}_{e}({\bf r},t)
=
\frac
{
  \hat H_{eg}({\bf r})
}
{\hbar\tilde\Delta}
\sum_{k=0}^\infty
\frac{(-i)^k}{\tilde\Delta^k}
\frac{\partial^k}{\partial t^k}
\hat \phi_g({\bf r},t)
\;.
\end{equation}
From the system of equations (\ref{nonlinear-g}),(\ref{nonlinear-e}),
with the center-of-mass motion neglected one can derive the relation
\begin{equation}
\frac{\partial^n}{\partial t^n}
\hat{\phi}_{g}({\bf r},t)
=
- i^n
\frac
{
  \hat H_{ge}({\bf r})
  \hat H_{eg}({\bf r})
}
{\hbar^2\tilde\Delta^{2-n}}
\sum_{k=n-1}^\infty
\frac{(-i)^k}{\tilde\Delta^k}
\frac{\partial^k}{\partial t^k}
\hat \phi_g({\bf r},t)
\;,
\quad
n=1,2,\dots
\end{equation}
Substituting this relation iteratively into equation (\ref{t-ind}),
we obtain
\begin{equation}
\label{series}
\hat{\phi}_{e}({\bf r},t)
=
\frac{\hat H_{eg}({\bf r})}{\hbar\tilde\Delta}
\sum_{m=0}^\infty
(-1)^m
a_m
\left[
     \frac{\hat H_{ge}({\bf r}) \hat H_{eg}({\bf r})}{\hbar^2\tilde\Delta^2}
\right]^m
\hat \phi_g({\bf r},t)
\;,
\end{equation}
\begin{eqnarray}
a_0
&=&
1,
\nonumber\\
a_{m+1}
&=&
\sum_{k_1=0}^{1}
\sum_{k_2=0}^{k_1+1}
\sum_{k_3=0}^{k_2+1}
\cdots
\sum_{k_m=0}^{k_{m-1}+1}
=
2\frac{(2m+1)!}{m!(m+2)!}
,
\quad
m=0,1,\dots
\end{eqnarray}
The zeroth order term ($m=0$) in equation (\ref{series}) corresponds
to the transition $|g\rangle \to |e\rangle$. The next term
($m=1$) corresponds to the transition
$|g\rangle \to |e\rangle \to |g\rangle \to |e\rangle$ and so on.
Therefore, different terms in equation (\ref{series}) describe multiple
transitions of the atoms between ground and excited states, caused
by the influence of the photons.
In a typical atom optical situation when the electromagnetic field
has a form of a standing wave these terms correspond to the processes
with the momentum transfer from the laser beam to the atoms
$\pm 2 \hbar {\bf K} (m+1)$, where ${\bf K}$ is a wave vector 
of the laser wave in a medium.

We assume that 
$\hat\epsilon=\hat H_{ge}\hat H_{eg}/(\hbar^2\tilde\Delta^2)$ 
acts only
on states $|\psi\rangle$ for which the series in equation (\ref{series})
converges, i.e., $|\psi\rangle$ can be decomposed into eigenstates
of $\hat\epsilon$ whose eigenvalues satisfy the condition 
$|\epsilon_n| < 1/4$. Then the result of the summation is given 
by
\begin{equation}
\label{e-g}
\hat\phi_{e}(t)
=
\frac{\hat H_{eg}}{\hbar\tilde\Delta}
\frac{\sqrt{1+4\hat\epsilon}-1}{2\hat\epsilon}
\hat\phi_g(t)
\;,
\end{equation}
and the equation for the ground state (\ref{nonlinear-g}) takes the form
\begin{equation}
\label{g-g}
i\hbar
\frac{\hat\phi_g}{\partial t}
=
\hat H_{cm} \hat\phi_g
+
\frac{\hbar\tilde\Delta}{2}
\left(
     \sqrt{1+4\hat\epsilon}-1
\right)
\hat\phi_g(t)
\;.
\end{equation}
Although we do not see explicitly on the r.h.s. of equations (\ref{e-g}), (\ref{g-g})
$\hat\phi_e$, it is still there, because $\hat{\bf E}^\pm_{loc}$ depends on $\hat\phi_e$
and this dependence, which is given by (\ref{def-e-local}), is very complicated.
Formally we could iteratively substitute equations (\ref{def-e-local}) and (\ref{e-g})
into (\ref{g-g}). However, by doing such a procedure
we would get an equation for the ground state wave function in a form which would be
impossible to use. Therefore, the algorithm described in this section does really
eliminate the excited state only in the case when the dipole-dipole interactions
are negligible, i.e., when $\hat {\bf E}^\pm_{loc} \approx {\bf E}^\pm_{in}$.
In order to eliminate
the excited state in the case when the dipole-dipole interactions play an important role
it is useful to employ in addition the Lorentz-Lorenz relation, 
which will be briefly discussed in the next section.

Because we are mainly interested in atom optical problems and want
to study the coherent evolution of the center-of-mass motion of the gas,
we shall neglect spontaneous emission. This is valid for situations
where the absolute values of the detunings are much bigger than
the spontaneous emission rates and Rabi frequencies 
$
\left|\Delta\right| \gg \gamma
\;,\;
\left|
     {\bf d}{\bf E}_{loc}^\pm/\hbar
\right|
$.
In order to do this approximation self-consistently we drop in 
the following the vacuum fluctuations
and the spontaneous emission rates $\gamma$ from our equations.
Then we may replace all the operators by macroscopic
functions.

\section{Local-field correction}

As it was mentioned in the previous section,
the solution of equations (\ref{def-e-local}), (\ref{e-g}), (\ref{g-g}) is
a rather complicated mathematical problem because these equations
contain explicitly the dipole-dipole interactions. In many particular situations
such a detailed microscopic description of matter is not necessary and it 
is more convenient to consider optical properties of the medium on a
macroscopic level. This can be done by introducing the macroscopic field
${\bf E}_{mac}({\bf r},t)$,
instead of the local field
${\bf E}_{loc}({\bf r},t)$ in the equations for the matter fields.

As in Ref.~\cite{GS} we can introduce the macroscopic field by imposing
the requirement that it is a solution of
the macroscopic Maxwell equations for a charge-free and current-free
polarization medium.
Using Maxwell equations and the definition of the local field
(\ref{def-e-local}), we get the following relation
\begin{equation}
\label{local-corr}
{\bf E}_{loc}^\pm({\bf r},t) 
=
{\bf E}_{mac}^\pm({\bf r},t) +
\frac{4\pi}{3}
{\bf P}^\pm({\bf r},t)
\;.
\end{equation}
This equation is often called in the literature the Lorentz-Lorenz relation. 
It constitutes the basis of the
local-field effects in 
classical, quantum and nonlinear optics.

\section{Nonlinear matter equation}

Let's consider first a special case of the low light intensity. In this case
we can keep only the first term ($m=0$) in equation (\ref{series}), which is
linear with respect to the electromagnetic field strength:
\begin{equation}
\label{adiabatic-series-l}
{\phi}_{e}({\bf r},t) 
=
-
\frac
{{\bf d} {\bf E}_{loc}^+({\bf r})}
{ 
  \hbar \Delta
}
{\phi}_{g}({\bf r},t)
\;.
\end{equation}
From equations (\ref{local-corr}), (\ref{adiabatic-series-l}) we obtain
\begin{equation}
\label{adiabatic-series}
{\phi}_{e}({\bf r},t) 
=
-
\frac
{{\Omega}^+({\bf r})}
{ 
  2 
  \Delta
}
\sum_{n=0}^\infty
  \left(
    \frac{4\pi}{3}
       \alpha
       \left|
  	   \phi_{g}({\bf r},t)
       \right|^2
  \right)^n
  \phi_{g}({\bf r},t)
\;,
\end{equation}
where the position dependent Rabi frequency
$
\Omega^+({\bf r})
=
2{\bf d} {\bf E}_{mac}^+({\bf r})/\hbar
$
is related to the macroscopic electric field,
$\alpha=-d^2/\hbar\Delta$ is the atomic polarizability.
The series in (\ref{adiabatic-series}) converges, provided that
$
    \frac{4\pi}{3}
       |\alpha|
       \left|
  	   \phi_{g}
       \right|^2
    <
    1   
$.
In this case we get
\begin{equation}
\label{adiabatic-sol}
{\phi}_{e}({\bf r},t) =
-
\frac
{{\Omega}^+({\bf r}) {\phi}_{g}({\bf r},t)}
{ 
  2 
  \Delta
  \left[
    1
    -
    \frac{4\pi}{3}
       \alpha
       \left|
  	   \phi_{g}({\bf r},t)
       \right|^2
  \right]
}
\;.
\end{equation}
This adiabatic solution has been obtained in Ref.~\cite{KBA99}
using slightly different technique. Note that a singularity
occurs in equation (\ref{adiabatic-sol}) under the condition
$
    \frac{4\pi}{3}
       |\alpha|
       \left|
  	   \phi_{g}
       \right|^2
    =
    1   
$.
However, as it follows from our present derivation, we never encounter
this singularity, because in the region
$
    \frac{4\pi}{3}
       |\alpha|
       \left|
  	   \phi_{g}
       \right|^2
    \ge
    1   
$
equation (\ref{adiabatic-sol}) is not valid. This was not clear from the derivation
given in Ref.~\cite{KBA99}.

Then substituting (\ref{adiabatic-sol}) in (\ref{nonlinear-g}),
we obtain as the result an
equation for the ground state matter field ${\phi}_{g}$~\cite{KBA99,KBA00}
\begin{equation}
\label{nonlinear-equation}
i\hbar
\frac{\partial \phi_{g}({\bf r},t)}{\partial t}
=
\left\{
   -
   \frac{\hbar^2\nabla^2}{2m}
   +
   \frac
   {\hbar \left|  \Omega^+({\bf r})\right|^2}
   {
     4
    \Delta
    \left[
      1
      -
      \frac{4\pi}{3}
         \alpha
         \left|
	     \phi_{g}({\bf r},t)
         \right|^2
    \right]^2
   }
\right\}
\phi_{g}({\bf r},t)
\;.
\end{equation}

Now we substitute the Lorentz-Lorenz relation (\ref{local-corr}) into
equations (\ref{e-g}),(\ref{g-g}) and keep the terms up to the order $1/\Delta^3$.
As a result we get the following equation for the ground-state wave function
\begin{equation}
\label{g3}
i\hbar
\frac{\partial \phi_{g}}{\partial t}
=
\hat H_{cm}
\phi_{g}
+
\frac
{\hbar \left| \Omega^+ \right|^2}
{4\Delta}
\left[
     1
     +
     \frac{8\pi}{3}
     \alpha
     \left|
          \phi_{g}
     \right|^2
     +
     3
     \left(
         \frac{4\pi}{3}
         \alpha
         \left|
              \phi_{g}
         \right|^2
     \right)^2
     -
     \frac
     {\left| \Omega^+ \right|^2}
     {4\Delta^2}
\right]
\phi_{g}
\;.
\end{equation}
Varying the density of atoms $\left| \phi_{g} \right|^2$, the light intensity,
which is proportional to $\left| \Omega^+ \right|^2$, and the magnitude and
the sign of the detuning $\Delta$, one can change
the effective potential for the ground-state matter field.

It is interesting to compare the nonlinear terms in equation (\ref{g3}) with
the terms in the equation for the condensate wave function, which takes
into account the effects of quantum fluctuations~\cite{QF}
\begin{equation}
\label{GPc}
i\hbar
\frac{\partial \phi_g}{\partial t}
=
\hat H_{cm}
\phi_g
+
\frac{4 \pi \hbar^2 a}{m}
\left[
     1
     +
     \frac{32}{3}
     \sqrt{\frac{a^3}{\pi}}
     \left|
         \phi_g
     \right|
\right]
\left|
    \phi_g
\right|^2
\phi_g
\;,
\end{equation}
where $a$ is a scattering length. 
The leading nonlinear terms in equations (\ref{g3}), (\ref{GPc}) are defined
by the quantities
\begin{displaymath}
U_E
=
-
\frac
{\left| \Omega^+ \right|^2}
{\Delta^2}
\frac{2\pi}{3}
d^2
\;,
\quad
U_C
=
\frac{4 \pi \hbar^2 a}{m}
\;,
\end{displaymath}
respectively. The typical orders of magnitude of the parameters for
alkali metal atoms are (in CGS system of units) $d \sim 10^{-18}\ {\rm esu}$,
$|a| \sim 10^{-7}\ {\rm cm}$, $m \sim 10^{-23}\ {\rm g}$, and we get the estimates
\begin{displaymath}
\left|
     U_E
\right|
\sim
\frac
{\left| \Omega^+ \right|^2}
{\Delta^2}
10^{-36}
\ {\rm erg}\;{\rm cm}^{-3}
\;,
\quad
\left|
     U_C
\right|
\sim
10^{-37}
\ {\rm erg}\;{\rm cm}^{-3}
\;.
\end{displaymath}
Therefore, we see that at $\left| \Omega^+ \right|^2 / \Delta^2 \sim 0.1$
$U_E$ and $U_C$ are of the same order of magnitude.

Higher order nonlinear terms in equations (\ref{g3}) and (\ref{GPc}) are
defined by
\begin{displaymath}
V_E
=
\frac
{\hbar \left| \Omega^+ \right|^2}
{4\Delta}
     3
     \left(
         \frac{4\pi}{3}
         \frac{d^2}{\hbar\Delta}
     \right)^2
         \left|
              \phi_{g}
         \right|^2
\;,\quad
V_C
=
\frac{4 \pi \hbar^2 a}{m}
     \frac{32}{3}
     \sqrt{\frac{a^3}{\pi}}
     \left|
         \phi_g
     \right|
\;,
\end{displaymath}
respectively.
For the densities $\left| \phi_{g} \right|^2 \sim 10^{14}\ {\rm cm}^{-3}$ 
and detunings
$\left| \Delta \right| \sim 10^8\ {\rm Hz}$, we have
\begin{displaymath}
\left|
     V_E
\right|
\sim
\frac
{\left| \Omega^+ \right|^2}
{\Delta^2}
10^{-38}
\ {\rm erg}\;{\rm cm}^{-3}
\;,\quad
\left|
     V_C
\right|
\sim
10^{-39}
\ {\rm erg}\;{\rm cm}^{-3}
\;.
\end{displaymath}
Thus, we see that the nonlinear terms in
equations (\ref{g3}) and (\ref{GPc}) have the same order of magnitude.

\section{Optical properties of the ultracold gas}

\subsection{Refractive index}

We substitute equation (\ref{e-g}) into the definition of the polarization field. 
This gives us a general nonlinear relation between the polarization
and the local field:
\begin{equation}
\label{P-gen}
{\bf P}^\pm
=
\alpha
\frac{\sqrt{1+4\epsilon}-1}{2\epsilon}
\left|
    \phi_g
\right|^2
{\bf E}_{loc}^\pm
\;.
\end{equation}

In the special case of the low light intensity we can use the adiabatic
solution (\ref{adiabatic-sol}). Then the expression for the polarization
takes the form
\begin{equation}
\label{P-lin}
{\bf P}^\pm
=
\alpha
\left|
    \phi_g
\right|^2
{\bf E}_{loc}^\pm
=
{\chi}
{{\bf E}}_{mac}^\pm
\;,
\end{equation}
where dielectric susceptibility is given by
\begin{equation}
\label{suscept-lin}
{\chi}
=
\frac
{
  \alpha
  \left|
      \phi_g
  \right|^2
}
{
     1
     -
     \frac{4\pi}{3}
     \alpha
     \left|
         \phi_g
     \right|^2
}
\;.
\end{equation}
This expression for the dielectric susceptibility $\chi$ leads to
the Clausius-Mossotti formula for the refractive index. As it follows
from our physical interpretation of the expansion (\ref{series}),
the Clausius-Mossotti formula corresponds to the quantum transition
of the type $| g \rangle \to | e \rangle$. Therefore, it takes into
account pair interactions between the atoms when one atom emits a
photon, then this photon is absorbed by another atom and so on.

If we keep the terms up to the order $1/\Delta^3$, the expression for the
polarization takes the form
\begin{equation}
\label{P-E_mac}
{\bf P}^\pm
=
\alpha
\left|
    \phi_g
\right|^2
\left[
     1
     -
     \frac
     {\left|{\bf d}{{\bf E}}_{loc}^+\right|^2}
     {\hbar^2 \Delta^2}
\right]
{\bf E}_{loc}^\pm
=
{\chi}
{{\bf E}}_{mac}^\pm
\;,
\end{equation}
where 
\begin{equation}
\label{def-suscept}
{\chi}
=
\alpha
\left|
    \phi_g
\right|^2
\left[
     1
     +
     \frac{4\pi}{3}
     \alpha
     \left|
         \phi_g
     \right|^2
     +
     \left(
         \frac{4\pi}{3}
         \alpha
         \left|
             \phi_g
         \right|^2
     \right)^2
     -
     \frac
     {\left|{\bf d}{{\bf E}}_{mac}^+\right|^2}
     {\hbar^2 \Delta^2}
\right]
\;.
\end{equation}

Dielectric susceptibility is a rather important parameter, because it
describes the propagation of the laser radiation inside a medium. In most
of the practical situations the electromagnetic processes are much faster
than the center-of-mass motion of the atoms. Therefore, $\chi$ can be
considered as a time-independent quantity.
Let us assume in addition that the spatial variations of the atomic density 
are not very large, such that $\nabla \chi \to 0$. Then 
$\mbox{div} \, {\bf E}_{mac}^\pm \approx 0$, and we have the following Helmholtz
equation for the macroscopic electric field
\begin{equation}
\label{maxwell-media}
\nabla^2 {\bf E}_{mac}^\pm  +
k_L^2 {n}^2 {\bf E}_{mac}^\pm = 0
\;,
\end{equation}
with the refractive index $n$ given by 
\begin{equation}
\label{n}
n
=
\sqrt{1 + 4 \pi \chi}
=
1
+
2\pi
\alpha
\left|
     \phi_g
\right|^2
\left[
    1
    +
    \frac{\pi}{3}
    \alpha
    \left|
 \phi_g
    \right|^2
    +
    \frac{5}{8}
    \left(
\frac{4\pi}{3}
\alpha
\left|
     \phi_g
\right|^2
    \right)^2
    -
    \frac
    {\left|{\bf d}{{\bf E}}_{mac}^+\right|^2}
    {\hbar^2 \Delta^2}
\right]
\;,
\end{equation}
which corresponds to the Kerr-type optical nonlinearity.

Equations (\ref{g3}), (\ref{maxwell-media}), (\ref{n})
can be considered as an atom optical analogue of the system of
Maxwell-Schr\"odinger equations used in quantum and nonlinear optics.
In general they have to be solved in a self-consistent 
way.

\subsection
{
  Polariton band gap
}

In this section we investigate the polariton spectrum of a condensate
interacting with an intense laser field. Let's consider a special case of
dilute atomic gas. In this case one can neglect the local-field correction
in equation (\ref{local-corr}) and we have
${\bf E}_{loc}^\pm({\bf r},t) \approx {\bf E}_{mac}^\pm({\bf r},t)$.
In order to simplify the analytical analysis we assume that the intensity
of the laser radiation is not too high. Then from Maxwell equations and
equation (\ref{P-gen}) we get the following expression for the refractive index
\begin{equation}
\label{n2-int}
n^2(\omega)
=
1-
\frac
{
  R
  \left(
       \omega-\omega_0
  \right)
}
{
  \left(
       \omega-\omega_0
  \right)^2
  +
  \left|\Omega^+\right|^2/4
}
\;,\;
R
=
  \frac{
        4\pi d^2
	\left|
	     \phi_g
	\right|^2
       }
       {\hbar}
\;,       
\end{equation}
which contains saturation effects, owing to the intense electric field.

The polariton spectrum $\omega(K)$ can be obtained from the equation
\begin{equation}
\label{disp}
K=\frac{\omega}{c}n(\omega)
\;.
\end{equation}
As it follows from equation (\ref{n2-int}) $n(\omega)$ is imaginary for the frequencies
$\omega$ within the range
\begin{equation}
\label{gap}
\omega_0 
+ 
\frac
{
  R
  -
  \sqrt
  {
    R^2
    -
    \left|\Omega^+\right|^2
  }
}
{2}
<
\omega
<
\omega_0 
+ 
\frac
{
  R
  +
  \sqrt
  {
    R^2
    -
    \left|\Omega^+\right|^2
  }
}
{2}
\;,\;
R > \left|\Omega^+\right|
\;.
\end{equation}
In this case equation (\ref{disp}) has no real solutions. This means that light
can not propagate in a condensate, i.e., we get a polariton band gap with the width
$\Delta_G = \sqrt{R^2 - \left|\Omega^+\right|^2}$. The gap width $\Delta_G$
increases with the density of the atoms in the ground state 
$\left|\phi_g\right|^2$ and decreases with the laser intensity.
In the case of the low light intensity ($\left|\Omega^+\right| \ll R$) we get
a well-known result $\Delta_G = R$~\cite{SS,Politzer}.
If $R \le \left|\Omega^+\right|$, equation (\ref{disp}) has only real solutions. This
means that in this case the polariton gap disappears and the medium is always
transparent.

Note that our analysis does not take into accout effects caused by the influence
of the spontaneous emission. As it was discussed in Ref.~\cite{Politzer},
the spontaneous emission smears out the polariton spectrum, providing
accessible polariton states inside the gap.

\section{Conclusion}

Starting from the microscopic model and making use of the
multipolar formulation of QED, we have derived the general system of 
Maxwell-Schr\"odinger equations for atomic creation and annihilation operators and
the propagation equation for the laser field. It describes
the modification of the properties of the external off-resonant laser
radiation in a medium due to dipole-dipole interactions and the influence
of this modification on the center-of-mass motion of the ultracold atoms
as a single dynamical process. The system can be used,
for instance, for the self-consistent analysis of linear and
nonlinear phenomena in atom optics at high
densities of the atomic system and high intensities of the laser radiation.

A general procedure of the elimination of the excited state has been developed.
The annihilation and creation operators of the excited state for large
detuning are
represented in the form  of a series expansion in powers of the inverse
detuning, which corresponds to multiple transitions between the ground
and excited electronic states of the atoms.

Optical properties of an interacting ultracold Bose gas are studied:
formula for the intensity-dependent
refractive index is derived and the polariton spectrum of a condensate
interacting with an intense laser field is investigated. 

\section*{Acknowledgments}

This work has been supported by the Deutsche For\-schungsgemeinschaft
and the Optikzentrum Konstanz. One of us (K.V.K.) would like to
thank also the Alexander-von-Humboldt Stiftung for financial support.
Helpful discussion with G.V.Shlyapnikov is gratefully acknowledged.

\newpage


\end{document}